\documentclass[b5paper,twoside]{jpconf}  %Default declaration%
\usepackage{graphicx,color, ulem}

\begin{document}

\title[Microwave Emission of Supra-arcade Structure]{Microwave Emission of Supra-arcade Structure associated with M1.6 Limb Flare}
\author[S. Kim, H. Bain, K. Shibasaki, and V. Reznikova]{Sujin Kim$^1$, Hazel Bain$^2$, Kiyoto Shibasaki$^1$, and Veronika Reznikova$^1$}
\address{$^1$Nobeyama Solar Radio Observatory, NAOJ,
Japan}
\address{$^2$Space Sciences Laboratory, University of California Berkeley, USA}
\ead{sjkim@nro.nao.ac.jp}

\begin{abstract}
We have investigated a supra-arcade structure, associated with an
M1.6 flare, which occurred on the south-east limb on 4th of
November 2010. It is observed in microwaves at 17 GHz with the
Nobeyama Radioheliograph (NoRH), soft X-rays in the range of 8-20
keV with the Reuven Ramaty High Energy Solar Spectroscopic Imager
(RHESSI), and EUV with the Atmospheric Imaging Assembly (AIA)
onboard the Solar Dynamics Observatory (SDO). As reported by
Reeves \& Golub (2011), the supra-arcade structure is observed
predominantly in the AIA 131 \AA\ channel, which contains a hot 11
MK component from Fe XIX (Boerner et al. 2011). While this hot
flare plasma lasts over the decay phase of the flare, it shows
some interesting characteristics in microwaves and soft X-rays: 1)
In the supra-arcade structure, the brightness temperature ($T_B$)
of the microwave emission increases gradually up to
2$\times$10$^4$ K, and 2) two soft X-ray sources appear: one
cospatial with the supra-arcade structure and another above the
post-flare arcade. We have derived the variation of emission
measure, density, and energy of the supra-arcade structure using
the $T_B$ obtained from 17 GHz microwave observations.
\end{abstract}

\section{Introduction}
Recently, Reeves \& Golub (2011) reported three events which show
hot plasma above the flare arcade observed by AIA/SDO. AIA
provides multi-wavelength EUV images with 1.2 $^{\prime\prime}$
resolution and 12 s temporal cadence. We have examined one of
these events, which shows significant microwave emission in the
supra-arcade structure, using NoRH, RHESSI, and AIA/SDO.

\section{Observations and Analysis}
We have examined the supra-arcade structure associated with an
M1.6 flare which occurred on the 4th of November 2010, on the
south-west limb. The GOES X-ray flux shows that the flare starts
at 23:52, peaks at 23:57, and gradually decreases before the start
of another flare at 00:50 on the 5th of November.

The flare arcade shows emission at all EUV wavelengths, while the
supra-arcade structure in the high corona is particularly
prominent at AIA 131 \AA\ (Fig. 1). Comparisons with other
wavelengths suggest T = 11~MK for the supra-arcade structure. The
supra-arcade structure is present in 131 \AA\ from the flare onset
until the onset of the subsequent flare on the 5th. The Hinode
X-ray Telescope (XRT) also observed this hot plasma. The $T_B$ of
the hot plasma in high corona at 17 GHz slowly increases and
extends out to the region of hot 131 \AA\ supra-arcade (blue
contours in Fig. 1). Red contours in Fig. 1 show RHESSI 8-20 keV
soft X-ray emission. The X-ray images were made using the 2-step
CLEAN method (Krucker et al. 2011). Two different sources we
identified: a bright compact source above the post-flare arcade;
and a faint, extended source cospatial with the supra-arcade
structure. Both sources appear continuously throughout the decay
phase.

Following Dulk (1985), for an optically thin, isothermal plasma,
the brightness temperature at a given frequency is related to the
plasma temperature $T$  (11 MK for our event) by
$T_B=T{\tau}_{\nu}$, where the optical depth
${\tau}_{\nu}={9.786\times{10}^{-3}}/{\nu^2}\ln\Lambda\
{EM}/{T^{3/2}}$. Using this we determine an emission measure $EM$
of 2$\times$10$^{31}$ cm$^{-5}$. The electron number density and
thermal energy were found to increase up to 5$\times$10$^9$
cm$^{-3}$ and 1$\times$10$^{30}$ erg, respectively.

\section{Results and Summary}
The supra-arcade thermal structure of an M1.6 flare is observed in
microwaves, EUV and soft X-rays. We found that the $T_B$ of the
supra-arcade at 17 GHz increases as the flare decays. A thermal
X-ray source is also present in the supra-arcade. The results show
unexpectedly high, and increasing, values of density and thermal
energy in supra-arcade flare plasma, persisting throughout a long
decay phase of around 50 min. To study the physical mechanism for
heating and injection of this plasma into the supra-arcade
structure, further work using AIA filter ratio temperature
estimation method and a more detailed investigation of the
morphological evolution of the hot plasma will be carried out.

\begin{figure}
\begin{center}
\includegraphics[width=12.5cm]{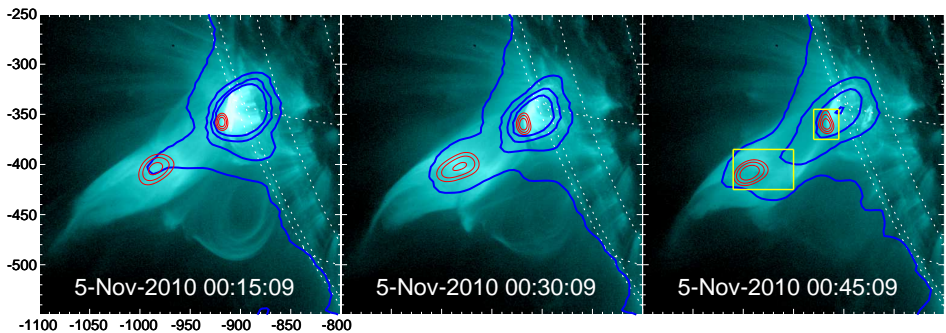}\vspace{-4ex}\end{center}
 \caption {\label{Figure 1} \small Time series
images taken by AIA 131 \AA\ with contours of NoRH 17 GHz (thick
blue line) and RHESSI 8-20 keV (thin red line) for the decay phase
of the flare. The levels of contour of NoRH 17 GHz are 1, 3, 5, 7
$\times$ 10$^4$ $T_B$. The levels of contour of RHESSI are given
in percentage of peak intensity ($I$): for bright source above
loop-top, 70, 80, 90 \% and for faint source on supra-arcade
structure, 3.3, 3.4, 3.5 \% (left) and 24, 25, 26 \% (middle), and
27, 28, 29 \% (right), respectively.\normalsize}
\end{figure}

\end{document}